\documentclass[11pt]{article}
\usepackage{epsf}
\newcommand{\beq}{\begin{equation}}
\newcommand{\eeq}{\end{equation}\noindent}
\newcommand{\bear}{\begin{eqnarray}}
\newcommand{\eear}{\end{eqnarray}\noindent}

\begin{document}

\begin{flushright}
BI-TH-97/41 \\
LPTHE Orsay 97/52 \\
October 1997
\end{flushright}

\begin{center}
\vspace{24pt}

{\Large \bf 4d Simplicial Quantum Gravity Interacting with \\
 Gauge Matter Fields}
\vspace{24pt}

{\large \sl S.~Bilke $^a$, Z.~Burda $^b$, A.~Krzywicki $^c$, 
 B.~Petersson $^a$, \\
 J.~Tabaczek $^a$ \,{\rm and}\, G.~Thorleifsson $^a$}
\vspace{10pt}

$^a$ Fakult\"{a}t f\"{u}r Physik, Universit\"{a}t Bielefeld, 
 33501 Bielefeld, Germany\\
$^b$ Institute of Physics, Jagellonian 
University, 30059 Krakow, Poland\\
$^c$ LPTHE, B\^{a}timent 211, Universit\'e Paris-Sud, 91405 Orsay, 
France\footnote{Laboratoire associ\'e au C.N.R.S.}
\vspace{10pt}

\begin{abstract}
The effect of coupling non-compact $U(1)$ gauge fields to
four dimensional simplicial quantum gravity 
is studied using strong coupling expansions and 
Monte Carlo simulations.
For one gauge field the back-reaction of the matter on the geometry
is weak. This changes, however, as more matter fields are introduced.
For more than two gauge fields the degeneracy of random
manifolds into branched polymers does not occur, and the branched
polymer phase seems to be replaced by a new phase with a negative 
string susceptibility exponent $\gamma$ and fractal dimension
$d_H \approx 4$. 
\end{abstract}
\end{center}
\vspace{15pt}

\section{Introduction and motivations}

\noindent
The statistical mechanics of random manifolds provides a possible
framework for a non-perturbative construction of a  quantum theory of 
gravity (there exist several excellent reviews, e.g.\ \cite{dav},
\cite{amb}). In this construction the standard recipes of
lattice field theory are adopted, with a notable extension: the 
lattice itself becomes a dynamical object instead of being an inert 
scaffold. For fixed topology, the summation over geometries 
involved in the partition function, approximating Feynman's path 
integral, is best implemented using the method of dynamical 
triangulations. The successes of this approach are particularly 
spectacular in two dimensions. The results obtained with models 
exactly solvable in the continuum formalism have been reproduced. 
Furthermore, completely new results, hardly attainable with 
another approach, have been derived. As an example, let us quote
the beautiful calculation of the "two-point" invariant correlation 
function in Ref.~\cite{kaw}. The method of dynamical triangulations 
is also applicable to four dimensional gravity. However, and this 
is not a surprise, the construction of a viable theory is much 
more difficult in four than in two dimensions.

At $d=2$ the Einstein term in the action is a topological
invariant and the dynamics is insensitive to the corresponding
coupling constant, as long as the topology of the manifold is kept
fixed. For $d > 2$ one finds that changing the (lattice analogue)
of the Newton constant can modify the intrinsic geometry of the
manifold. A phase transition is observed, separating two phases: 
in one the manifolds are crumpled, in the other they are elongated 
and resemble branched polymers. The transition was first 
discovered at $d=3$, and is there of first order \cite{bou}. 
The latent heat is large, in agreement with the generic picture 
suggested by mean field arguments \cite{bia}. The transition 
found at $d=4$ was initially thought to be continuous \cite{amb2}, 
but a more recent study \cite{bia2} has furnished the evidence that 
it is actually of first order too, but with a very small latent 
heat. This finding has been confirmed by further work \cite{deb}. 
The  smallness of the latent heat remains, however, a mystery.
On the other hand, the dynamics of the transition has been to
a large extent elucidated \cite{hot,bbpt,cat}.

In Ref.~\cite{jur} it has been emphasized that the crumpled $\to$
elongated transition occurs in all dimensions, but at $d=2$ this
happens only when the central charge $c$ of matter fields is large
enough. By lowering $c$ below unity one enters 
the Liouville phase, where a sensible continuum limit can
be defined. There exists a heuristic argument \cite{cates}
explaining that: when, increasing $c$, one crosses the magic
value $c=1$ there occurs a condensation of metric singularities
(``spikes'') and a collapse of the Euclidean space-time. The
continuum formalism and the known dynamics of the conformal 
factor (Liouville action) can be used to calculate
the free energy of spikes. The authors of Ref.~\cite{jur} have
suggested that a similar phenomenon might occur at $d=4$.
Using the effective action for the conformal factor calculated in
Ref.~\cite{ant}, the free 
energy/spike is\footnote{The relevant spike configuration is 
$\phi = \ln{\{1 + \rho^2/\left[(x-x_0)^2 +a^2\right]\}}$. 
The metric is proportional to $e^\phi$ and 
$a \ll \rho < \mbox{\rm const}\times a$.}:
\beq
F = \left[ {{1411+N_S+11N_F+62N_V-28} \over {360}} - 4 \right] 
\; \ln{{\rho} \over a} \;.
\label{fener}
\eeq
The logarithmic factor on the right-hand side can be made
arbitrarily large by taking a sufficiently spiky spike. The
symbols $N_{S,F,V}$ refer to the number of massless scalar,
fermion and vector fields contributing to the trace anomaly.
The number $1411$ comes from a one loop calculation of
the contribution of transverse gravitons and $28$
from ghosts and from quantum fluctuations of the conformal
factor itself.
This suggests that in the absence of
matter fields the theory has no sensible vacuum and that this
instability is lifted when the number of matter fields is
large enough. As emphasized in Ref.~\cite{ant}, it is remarkable
that the sign of the contributions of matter and ghost fields is 
in 4d opposite to that found in 2d. Thus in 4d one stabilizes the 
theory by {\em adding} matter fields, while in 2d the contrary 
happens. This idea has been further expanded in Ref.~\cite{ant2}.

In two dimensions the effective action behind the estimate of the
contribution of spikes can be derived essentially rigorously.
It should be emphasized that this is not the case in four dimensions,
because it is not clear how to treat the transverse part of
the metric.
Nevertheless, we consider the conjecture discussed above 
sufficiently interesting to
investigate the influence of matter on the geometry in 
4d simplicial quantum gravity, although earlier investigations 
have not observed any non-trivial effects of 
the matter sector \cite{matter}. 

Since the right-hand side of Eq.~(\ref{fener}) 
depends most strongly on
the number of vector fields, we consider 
in this paper only continuous
vector gauge fields, which have not been studied previously
in 4d simplicial gravity. 
The model is defined in Section~2. In Section~3 we discuss
the strong coupling expansion of the partition 
function for simplicial
gravity interacting with any number of gauge 
vector fields. Our analysis
of this series indicates that polymerization is 
suppressed when $N_V > 2$.
We check this point in Section~4, where the 
results of our Monte Carlo 
simulations are presented. Our speculations and conclusions
form the content of Section~5.

\section{The model}

\noindent
The action is a sum of two parts. The first is the Einstein-Hilbert
action, which for a 4d simplicial manifold reads:
\beq
S_G \,=\, -\kappa_2 N_2 + \kappa_4 N_4 \;,
\eeq
where $N_k$ denotes the number of $k$-simplexes. The second part is
\beq
S_M \,=\, \sum_{t_{abc}} o(t_{abc}) \left[ A(l_{ab}) + 
A(l_{bc}) + A(l_{ca})\right]^2 \;,
\label{sm}
\eeq
where $A(l_{ab})$ is a $U(1)$ gauge field 
living on link $l_{ab}$ and
$A(l_{ab}) = - A(l_{ba})$. The sum extends 
over all triangles $t_{abc}$
of the random lattice and $o(t_{abc})$ denotes 
the order of the triangle
$t_{abc}$, i.e.\ the number of simplexes sharing this triangle. 
Since we adopt a non-compact version of the theory,  with a
Gaussian action, there is no need to introduce 
a coupling in front of the sum on the right-hand side above.

We work in a pseudo-canonical ensemble of (spherical) manifolds, 
with almost fixed $N_4$. The model is 
defined by the partition function:
\beq
Z(\kappa_2, \bar{N}_4) \,=\,  \sum_{T} W(T) \int' 
\prod_{l \in T} dA(l) \; \; {\rm e}^{\textstyle - S_G - S_M - 
{\delta \over 2} (N_4- \bar{N}_4)^2 } .
\label{part}
\eeq
The sum is over all distinct triangulations $T$ 
and $W(T)$ is the symmetry factor taking 
care of equivalent re-labelings of vertexes. The prime indicates
that the zero modes of the gauge field are not 
integrated. As is well known,
the volume conserving local move is not ergodic, hence we must
allow the volume to fluctuate. 
The quadratic potential term added to the action ensures,
for an appropriate choice of $\delta$, 
that these fluctuations are small. 
The parameter $\kappa_4$ is adjusted so that $N_4$ fluctuates
around a mean volume $\bar{N}_4$. 

The Monte Carlo algorithm is constructed 
following one of the standard
recipes of grand-canonical simulations 
\cite{jkp}. A geometry move is
done in two steps: first one integrates 
out the changing matter degrees
of freedom, then the move is accepted/rejected based on a 
Metropolis test. If it is accepted, new matter fields are
generated from a heatbath distribution, if new links are created, 
or disregarded if links have been removed. 
A geometry sweep consists of $N_4$
attempted (randomly chosen) geometry moves. Matter fields
are updated using heatbath and overrelaxation algorithms. 
In a heatbath update the field is 
generated from the Gaussian distribution 
appearing
in the partition function. In an overrelaxation update the field is
given its ``image'' value: $A(l) \to - A(l) + 2\bar{A}$, where
$\bar{A}$ is the average over neighboring fields. A heatbath,
or overrelaxation, sweep consists of $N_1$ attempted updates. 
The relative frequency of geometry and matter sweeps is chosen so
as to minimize both the auto-correlation time and the CPU demand; 
we use typically  one heatbath sweep followed by 
one overrelaxation sweep, after each geometry sweep. 
Measurements are done when $N_4=\bar{N}_4$, and are 
separated by a constant number of ``passes'' through $N_4 =\bar{N}_4$.
This number is chosen so that the system makes approximately 10 
geometry sweeps between two successive measures.

\section{The strong coupling expansion}

\noindent
In parallel to the development of the Monte Carlo code, we have
calculated successive terms of the strong coupling expansion of
the partition function Eq.~(\ref{part}). 
This calculation is split into two distinct parts:
First, for a given triangulation,
one has to find the symmetry factor: 
$W(T) = 1/(\mbox{\rm number of equivalent 
re-labelings of} \; T)  \; $.
Second, one has to calculate the determinant $\Delta$, resulting 
from the Gaussian integration over one species of gauge fields 
(with unobservable zero modes kept fixed).

We found the weights up to $N_4 = 18$ by inspection,
in order to test our codes and to get some insight into the problem. 
For larger manifolds we used pure gravity Monte Carlo simulations to 
identify the distinct triangulations  
and to determine the corresponding symmetry factors
numerically, counting the relative frequency of the triangulations.
The symmetry factors are also calculated explicitly, by
going through all permutations of vertex labelings; comparing this
to the Monte Carlo results serves as a consistency check on the 
identification.
Finally, the determinant $\Delta$ is calculated using Maple. 
We have pushed the calculation 
up to $N_4=30$; the different contributions to the
partition function, Eq.~(\ref{part}), are 
shown in Table~1 for zero, one
and three vector fields.  

\begin{table}
{\small
\caption{\small The number of different graphs, $N_g$, for a fixed 
 volume $N_4$ and fixed number of vertexes $N_0$,
 and the corresponding weights $W_{N_V}(N_4,N_0)$.  
 This is shown both for pure gravity, and one and three vector fields
 coupled to gravity ($N_V = 0$, 1, and 3).  All weights are 
 normalized with the value at $N_4 = 6$.}
\vspace{10pt}
\begin{center}
\begin{tabular}{|rr|rrll|} \hline
$N_4$ & $N_0$  & $N_g$ & $W_0$  & $W_1$ & $W_3$\\  \hline
6  &  6 &    1 &     1    & 1  & 1 \\
10 &  7 &    1 &     3    & $0.097638467\ldots$ 
                          & $1.03423\ldots\times10^{-4}$ \\
12 &  7 &    1 &     5    & $0.030058406\ldots$ 
                          & $1.08632\ldots\times10^{-6}$ \\
14 &  8 &    1 &    15    & $0.018550484\ldots$ 
                          & $2.83715\ldots\times10^{-8}$ \\
16 &  8 &    2 &   255/4  & $0.015777808\ldots$ 
                          & $9.68553\ldots\times10^{-10}$ \\
18 &  8 &    3 &   110    & $0.005500465\ldots$ 
                          & $1.38182\ldots\times10^{-11}$ \\
   &  9 &    3 &    95    & $0.004759295\ldots$ 
                          & $1.19996\ldots\times10^{-11}$ \\
20 &  8 &    2 &   225    & $0.002512817\ldots$ 
                          & $3.15034\ldots\times10^{-13}$ \\
   &  9 &    7 &   693    & $0.007291315\ldots$ 
                          & $8.14758\ldots\times10^{-13}$ \\
22 &  9 &   15 &  2460    & $0.005728290\ldots$ 
                          & $3.27573\ldots\times10^{-14}$ \\
   & 10 &    7 &   690    & $0.001447804\ldots$ 
                          & $6.46761\ldots\times10^{-15}$ \\
24 &  9 &   13 & 16365/2  & $0.004226212\ldots$ 
                          & $1.17586\ldots\times10^{-15}$ \\
   & 10 &   34 & 14625/2  & $0.003378959\ldots$ 
                          & $7.45244\ldots\times10^{-16}$ \\
26 &  9 &   50 & 17865    & $0.001946262\ldots$ 
                          & $2.34367\ldots\times10^{-17}$ \\
   & 10 &  124 & 39645    & $0.003936950\ldots$ 
                          & $3.97116\ldots\times10^{-17}$ \\
   & 11 &   30 &  5481    & $0.000491334\ldots$ 
                          & $4.06700\ldots\times10^{-18}$ \\
28 &  9 &   89 & 291555/7 & $0.001058334\ldots$ 
                          & $6.96159\ldots\times10^{-19}$ \\
   & 10 &  415 & 182820   & $0.004119603\ldots$ 
                          & $2.16619\ldots\times10^{-18}$ \\
   & 11 &  217 &  77057   & $0.001534637\ldots$ 
                          & $6.33995\ldots\times10^{-19}$ \\
30 &  9 &  139 &  73860   & $0.000457581\ldots$ 
                          & $1.78973\ldots\times10^{-20}$ \\
   & 10 & 1276 & 672821   & $0.003427165\ldots$ 
                          & $9.13722\ldots\times10^{-20}$ \\
   & 11 & 1208 & 564000   & $0.002507268\ldots$ 
                          & $5.17157\ldots\times10^{-20}$ \\
   & 12 &  143 &  46376   & $0.000179907\ldots$ 
                          & $2.84432\ldots\times10^{-21}$ \\  \hline
\end{tabular}
\end{center}
}
\end{table}

The big advantage of the strong coupling approach 
is that once the weights 
and the determinants have been calculated, it does not cost any effort 
to change $\kappa_2$ and/or the number of matter fields. The series
has been analyzed using the ratio method to extract the string
susceptibility exponent $\gamma$ (cf.\ Ref.~\cite{dav2}), assuming the
asymptotic behavior of the partition function:
\begin{equation}
Z(\bar{N}_4) \;\sim\; {\rm e}^{\textstyle \kappa_4^c \bar{N}_4}
 \; \bar{N}_4^{\gamma - 3}. 
\label{simgam}
\end{equation}
Actually, only every second
term of the series can be used for this purpose. One finds an
oscillation in the behavior of the coefficients; a closer look
reveals that the complete series is a sum of two. If one estimates 
the critical coupling using these series separately, the two estimates
approach the correct value from below and from above, respectively.
For $\gamma$ we obtained more reliable 
results using the terms corresponding 
to $N_4=6, 10, 14, \dots$\footnote{A related regularity is observed in 
the distribution of the volumes of minimal neck baby universes
(minbu's) on the manifold. It has two branches, each corresponding to
every second minbu size, which merge together at large minbu
volumes.  Also in this case, the series $N_4=6,10,14,\ldots$ 
is less affected by finite size effects.}

\begin{figure}
\epsfxsize=4in \centerline{ \epsfbox{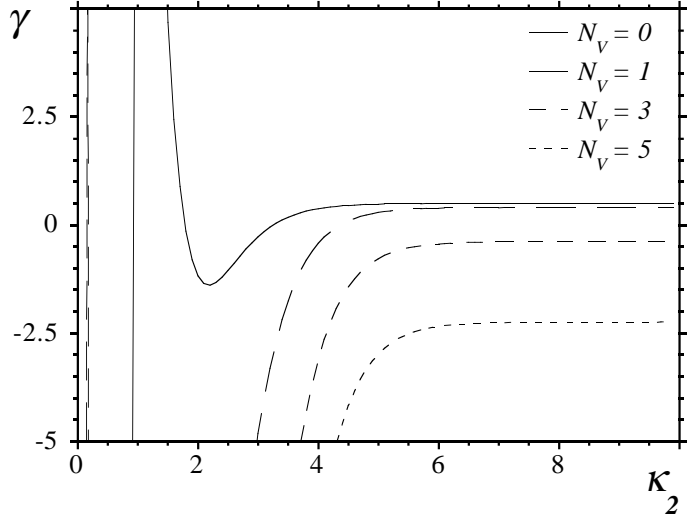}}
\caption[fig1]{\small Variations of $\gamma$ with
 $\kappa_2$, for 0, 1, 3, and 5 vector fields coupled
 to gravity.  These values are obtains using the ratio
 method to analyze the strong coupling series, including terms  
 corresponding to $N_4 = 6$, 10, 14, 18, 22, 26, and 30.}
\label{fig1}
\end{figure}

Our results can be summarized as follows:
\begin{itemize}
\item[({\it i})] 
For fixed (large) $\kappa_2$ the susceptibility exponent 
$\gamma$ becomes negative when the number $N_V$ of vector fields 
increases.
\item[({\it ii})] 
For fixed $N_V$ and varying $\kappa_2$, the exponent $\gamma$
has a characteristic behavior: it is large negative for small enough 
$\kappa_2$, while it tends to a constant 
value for large $\kappa_2$. This
limiting value is ${1 \over 2}$, as expected, 
for pure gravity, and becomes
negative for $N_V > 2$. For intermediate 
$\kappa_2$ the estimate of the
exponent is highly unstable, which is presumably a reflection of the
existence of a phase transition. 
This is illustrated in Fig.~(\ref{fig1}).
For $\kappa_2 > \kappa_2^c$, on the other hand, the results are
very stable, both with respect to variations in $\kappa_2$
and number of terms included in the analysis.
\end{itemize}

These results suggest that for $N_V > 2$ the branched polymer
phase disappears, and is replaced by a {\it new}
phase with a negative $\gamma$.
This is similar to what happens in two dimensions, except in
four dimensions $\gamma$ becomes more negative as the number 
of vector fields is increased.  All this is in qualitative
agreement with the expectations described in the Introduction.
As these results have been obtained using relatively
small triangulations, we have also 
carried out a series of Monte Carlo simulations to substantiate
this picture; those results
are presented in the next section.

\section{Monte Carlo simulations}

To understand better the phase structure of the
model, as the number of vector fields $N_V$ is increased, 
we have performed Monte Carlo simulations using one
and three vector fields, mostly on lattices with no more than
16K simplexes.  As these are rather modest lattice
volumes, we present our results with the appropriate
reservations; it is well known that for pure gravity,
simulations with such small volumes can give misleading
informations, especially about the nature of the phase
transition.  We nevertheless believe that the results of
our simulations, which agree well with the strong coupling 
expansion, present the correct qualitative picture.  

As we want to compare our results to the corresponding simulations 
of pure gravity, we summarize what is observed in that case:
There is a first-order transition \cite{bia2,deb} separating
a small $\kappa_2$ (strong coupling) crumpled phase from
a large $\kappa_2$ (weak coupling) branched polymer phase;  
$\kappa_2^c \approx 1.29$.  The crumpled phase is characterized by
{\it two} singular vertexes, connected to an extensive fraction
of the total volume, i.e.\ their local volume grows like $N_4$.
Those vertexes are joined together by a sub-singular link; its local
volume grows like $N_4^{2/3}$.  This singular structure is not
present in the branched polymer phase, it dissolves at the phase
transition.  In simulations on small lattices one actually 
observes that the transition occurs in two steps \cite{bbpt,cat,tab};
first the sub-singular link is dissolved, later, at weaker coupling,
the singular vertexes disappear.  Those two sub-transitions,
however, seem to merge as the volume is increased.

\begin{figure}
\epsfxsize=4in \centerline{ \epsfbox{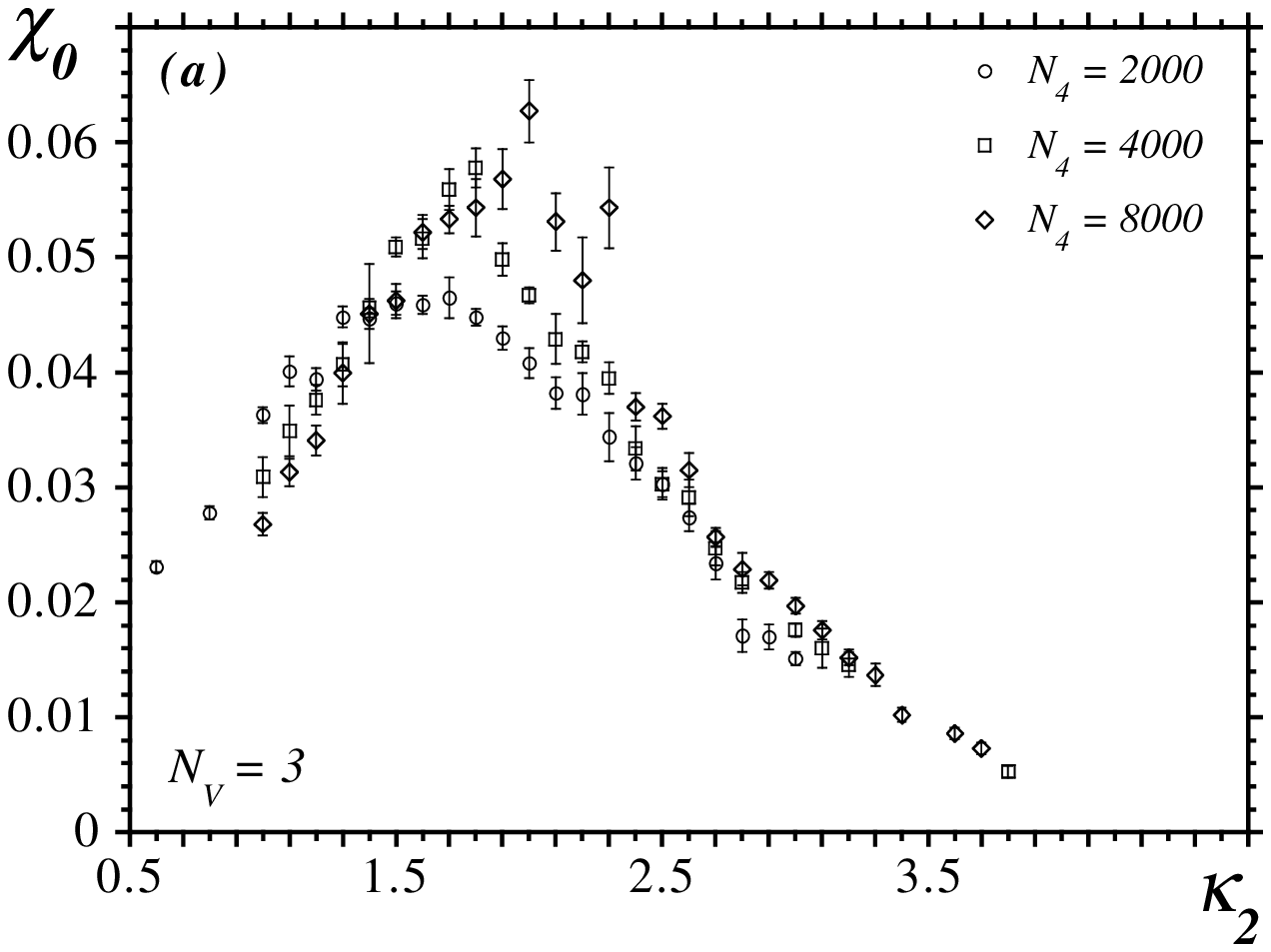}}
\epsfxsize=4in \centerline{ \epsfbox{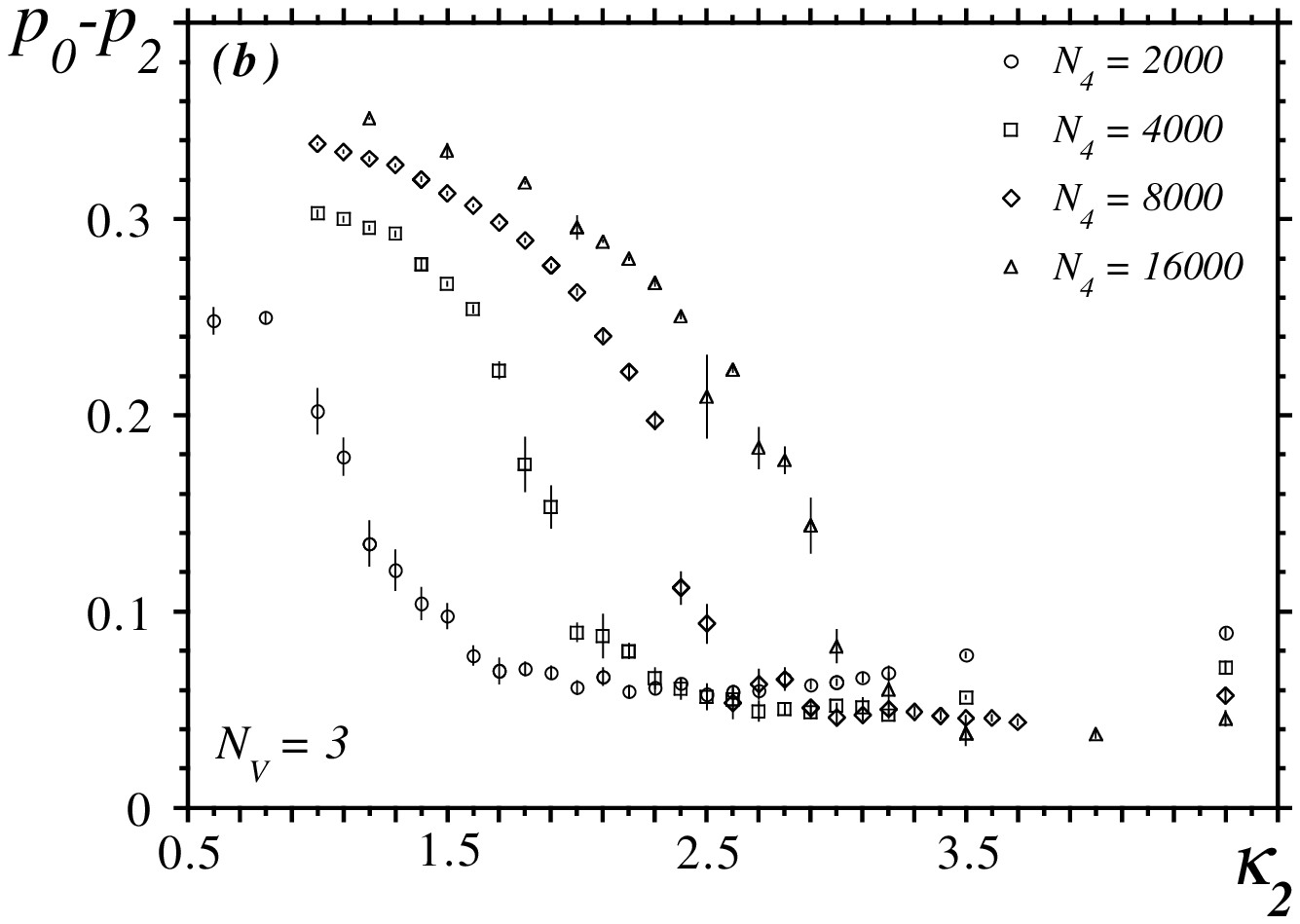}}
\caption[fig2]{\small {\it (a)} The node susceptibility $\chi_0$,
 for three vector fields coupled to gravity, {\it vs}.\ $\kappa_2$.
 For $N_4 = 16000$ the statistics is not sufficient for a
 reliable estimate of $\chi_0$.
 {\it (b)} The corresponding change in the difference between the
 orders of the first and of the third most singular 
vertex, $p_0$ and $p_2$,
 normalized with the total volume $N_4$.} 
\label{fig2}
\end{figure}

This picture does not change much if one adds one vector
field to the model.  We still observe two peaks in the
node susceptibility,
$\chi_0 = (\left<N_0^2\right> - \left<N_0\right>^2)/N_4$,
presumably associated with the two sub-transitions discussed above.
Those transition separate the usual crumpled and
branched polymer phases.
The second transition, at larger $\kappa_2$, is more pronounced,
and for the largest volume (32K) we observe
a clear signal of a first-order transition, i.e.\
fluctuations between two states in the timeseries of the energy.
The corresponding latent heat appears to be
smaller than in the case of pure gravity --- maybe an indication that
the transition becomes softer as matter is added.

A dramatic change occurs, however, when we introduce three
vector fields to the model.  We only observe a single
peak in the node susceptibility, at $\kappa_2 \approx 2$. 
Its height increases with the volume; see Fig.~(\ref{fig2}{\it a}).
Presently, our data are not good enough to decide whether 
this is a real phase transition and, in the affirmative,
what is the nature of the transition and the corresponding
critical coupling. 
Simulations with higher statistics and at larger volumes
are needed for that. We also observe a change in the geometry of the
manifolds coinciding roughly with the peak in $\chi_0$: 
at strong couplings the system is in the crumpled phase, one has
two well identified singular vertexes of almost the same order, 
separated by a large gap from the orders of other vertexes. As
$\kappa_2$ increases the orders of these vertexes merge 
with the rest of the vertex order distribution.
This is shown in Fig.~(\ref{fig2}{\it b}), 
where is plotted the difference
between the orders of the first and the third most 
singular vertex: $p_0 - p_2$.  
Combined, these results indicate that there still
is a phase transition in this model.  

\begin{table}
\caption{\small Measured values of the string susceptibility
 exponent $\gamma$ in the weak coupling
 (large $\kappa_2$) phase, for $N_V = 3$.}
\begin{center}
{\small
\begin{tabular}{|c|c|c|} \hline
 $N_4$  &  $\kappa_2 = 4.5$  & $\kappa_2 = 6.0$ \\  \hline
 2000   &  -0.22(2) & \\
 4000   &  -0.18(3) &  -0.17(4) \\
 8000   &  -0.23(3) & \\
 16000  &  -0.30(6)   &  -0.12(6)  \\  \hline
\end{tabular}
}
\end{center}
\end{table}

\begin{figure}
\epsfxsize=4in \centerline{ \epsfbox{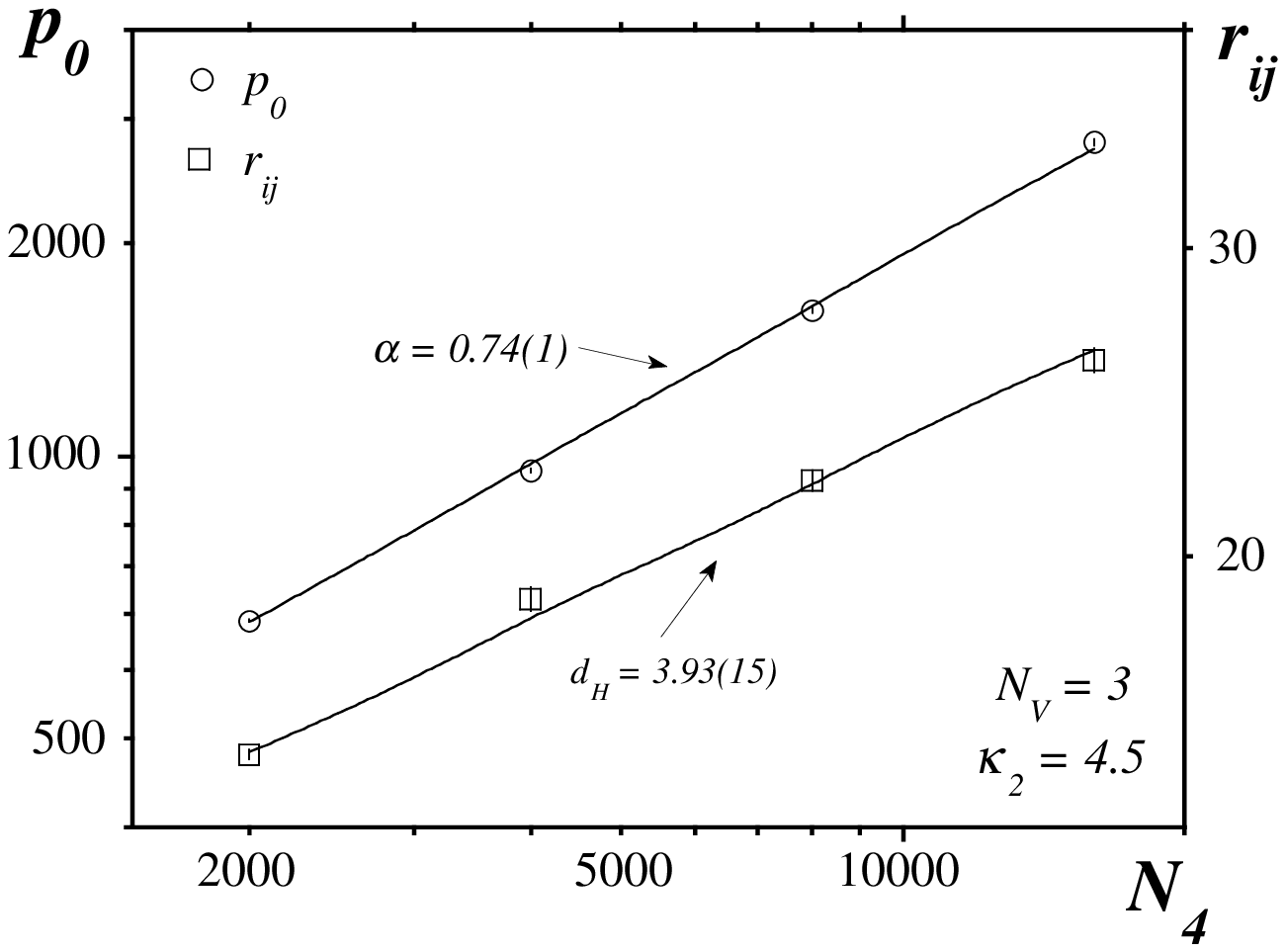}}
\caption[fig3]{\small The scaling of both the largest vertex order,
 $p_0$, and of the average distance between two simplexes,
 $r_{ij}$, in the new weak coupling phase, for $N_V = 3$ and
 at $\kappa_2 = 4.5$.  The lines are the best linear fits
 to the data.}
\label{fig3}
\end{figure}

What about the respective phases?
At small $\kappa_2$ there is the usual crumpled phase, but the
weak coupling phase is no longer that of branched polymers. 
As mentioned above, it does not contain a singular vertex.
This is best demonstrated by the scaling
of the largest vertex order, which we show in 
Fig.~(\ref{fig3}) for $\kappa_2 = 4.5$.
From the slope we see that the local volume grows
like $N_4^{3/4}$, i.e.\ it is a vanishing fraction of the
total volume.  
We have also measured the
string susceptibility exponent $\gamma$ in this phase,
extracted from the minbu distribution \cite{baby}.
This we did for two values of 
$\kappa_2$ (4.5 and 6.0) and at different
volumes (see Table~2).
In all cases we get a consistent negative value,
$\gamma \approx -0.2$, in reasonable 
agreement with the prediction from
the strong coupling expansion, which is $\gamma \approx -0.38$. 
The fits to the minbu distributions
are very good, which confirms the validity of the assumed asymptotic 
behavior Eq.~(\ref{simgam}). 
In contrast, it is not possible
to extract any reliable value of $\gamma$ in the crumpled phase,
nor close to the phase transition.  Finally, we have 
measured the Hausdorff or fractal dimension $d_H$, using 
the scaling of the average distance between the simplexes   
\cite{haus}:
\begin{equation}
\left < r_{ij} \right >_{N_4} \;=\; 
\left<\sum_{r=0}^{\infty}  r \;n(r) \right> 
\;\sim\; N_4^{1/d_H} \;,
\end{equation}
were $n(r)$ counts the number of simplexes at a 
geodesic distance $r$ from a marked simplex.  
This is included in Fig.~(\ref{fig3}); a fit to Eq.~(\ref{fig3})
gives $d_H = 3.97(15)$.

One apparent pathology of this new phase, is that the node number
is very close to its upper kinematic bound,
i.e.\ $\langle N_0\rangle/N_4 \approx 0.25$. What 
this implies for the nature
of the phase is not yet clear to us.

\section{Summary and discussion}

\noindent
The most important result of this work is the discovery that by
introducing matter fields one can prevent the standard collapse 
of 4d random manifolds into branched polymers. Contrary to all
earlier investigations \cite{matter}, we find a strong back-reaction 
of matter on geometry. This opens new research possibilities
although, of course, one still faces 
many uncertainties and pathologies.

As already mentioned, the (discontinuous) 
phase transition observed
with $N_V=1$ seems softer than 
that in pure gravity. Eq.~(\ref{fener})
indicates that a weighted sum of $N_{S,F,V}$ controls the behavior
of the theory; it is not excluded that, for some appropriate
``combination'' of matter fields, the transition might 
become continuous. This would be very exciting: 
one would  have a relation between the matter content of
the theory and its very existence.  

The nature of the new weak coupling phase, which 
we have observed, for $N_V = 3$, is also very intriguing. 
Our results suggest that this phase is characterized by a non-trivial
negative $\gamma$ and that it has fractal dimension four,
the same as the flat space.  Yet it has some pathologies.
Although its largest vertex order scales sub-linearly with
the volume, it grows much faster than, for example, the largest 
vertex order in pure 2d gravity (where the growth is logarithmic).
In addition, we observe that the number of nodes of the manifolds 
is very close to its kinematic bound.
In spite of this, it is not excluded that a
non-trivial continuum limit can be taken in this phase, 
i.e.\ that the whole phase is critical, as a negative
susceptibility exponent would suggest.  That would be
analogous to what happens in two dimensions, but  
it is not clear what would be the nature of
such a continuum theory, obtained without tuning
the Newton's constant to a critical value.

\section*{Acknowledgments}
We are indebted to B. Klosowicz for help.
We have used the computer facilities of the CRI at Orsay 
and of the CNRS computer center IDRIS, the HRZ, Univ.\
Bielefeld and HRZ Juelich. S.B.\ and J.T.\ were supported
by the DFG, under the contract PE340/3-3, and  
G.T.\ by the Humboldt Foundation. Z.B.\ has benefited from
the KBN grant no. 2 P03B 044 12.

\end{document}